\documentstyle[prl,aps,psfig]{revtex}
\tighten
\begin{document}
\draft
\twocolumn

\title{Superconductivity-Induced Anderson Localisation.}

\author{D.E. Katsanos, S.N. Evangelou and  C.J. Lambert$^\dagger$}
\address{ Department of Physics, University of Ioannina, Ioannina 45 110,
Greece}
\address{ $^\dagger$
School of Physics and Chemistry,
Lancaster University, Lancaster LA1 4YB, U.K.}

\date{\today}
\maketitle

\begin{abstract}

We have studied the effect of a random superconducting order
parameter on the localization of quasi-particles,
by numerical finite size scaling of
the Bogoliubov-de Gennes tight-binding Hamiltonian. 
Anderson localization is
obtained in $d=$ 2 and a mobility edge where the states 
localize is observed in $d=$ 3. The critical
behavior and localization exponent are universal
within error bars both for real and complex random order parameter.
Experimentally these results  imply a suppression of
the electronic contribution to thermal transport from states
above the bulk energy gap.

\end{abstract}


During the past few years phase-coherent transport in hybrid superconducting
structures has emerged as a new field of study, bringing together
the hitherto separate areas of superconductivity and mesoscopic physics.
Recent experiments have revealed a variety of unexpected phenomena,
including zero-bias anomalies \cite{1,2,3}, 
re-entrant and long-range behaviour \cite{4} and
phase-periodic transport \cite{5,6,7,8,9}.
These experiments can all be described
by combining traditional quasi-classical Green's function techniques
with boundary conditions
derived initially by Zaitsev \cite{10} and simplified by Kuprianov and
Lukichev \cite{11} or alternatively by generalised current-voltage relations 
\cite{12} based on a  multiple scattering approach to phase-coherent
transport. The latter approach focusses attention to Andreev scattering 
\cite{13}, whereby an electron can coherently evolve into a hole and
vice versa, without phase breaking.

The aim if this Letter is to address a new phenomenon, not describable by
quasi-classical techniques, namely
the onset of quasi-particle Anderson localisation due to spatial
fluctuations in a superconducting order parameter. In contrast with
all of the above experiments, where the superconducting
order parameter $\Delta(\underline r)$ is typically
homogeneous, there are many situations in which $\Delta(\underline r)$
varies randomly in space, even though the underlying
normal potential is perfectly ordered. One example is provided by the melting
of a flux lattice \cite{14,15}
in an otherwise perfectly crystalline high $T_c$
superconductor. Another should occur in anisotropic superconductors, where
by analogy with $^3$He-A, disordered textures can arise when an anisotropic
phase is nucleated from a more symmetric phase such as $^3$He-B.
In the first of these examples, the order
parameter is not quenched. Nevertheless, close to the melting curve,
the time scale for changes in $\Delta(\underline r)$
can be made arbitrarily long and therefore in the spirit of the
Born - Oppenheimer approximation, it is reasonable to freeze the disorder and
when necessary, treat any temporal fluctuations as a contribution 
to the inelastic scattering lifetime.

In one dimension, it is straightforward to
demonstrate \cite{16} that fluctuations in $\Delta(\underline r)$
alone can localise the excitations, even at energies
high above the bulk energy gap. However,
localisation in strictly one-dimension is
of little interest experimentally and therefore
in this Letter, we address the
question of whether or not  superconductivity induced
Anderson localisation occurs in higher dimensions. 
Early analytic work by \cite{17,18},
suggested that in the presence
of time reversal symmetry, states of energy $E=0$ are localised for
dimensions $d \le 2$, while in the absence of time reversal symmetry such
states are localised in all dimensions. 
However calculations using a numerical finite size scaling approach
\cite{19}
were inconclusive and to date there has been no experimental confirmation of
these predictions.
In this Letter we provide the first firm numerical evidence  for
superconductivity induced Anderson localisation in $d=$ 2 and 3 
dimensions and for the first time
compute the exponent $\nu$ controlling the divergence 
of the localisation length $\xi$ at the mobility edge in $d=$ 3.

To address the question of superconductivity induced Anderson localisation,
we analyze the tight-binding Bogoliubov-de Gennes  equations
\begin{equation}\matrix{ E\psi_i(E)
=&\epsilon_i \psi_{i}(E)
-\gamma\sum_{j}    \psi_{j}(E)
+  \Delta_{i} \phi_{i}(E),
\cr \cr
E\phi_i(E) =&-
\epsilon_i \phi_{i}(E)+\gamma^*\sum_{j}    \phi_{j}(E)
+ \Delta^*_{i}\psi_{i}(E) }\label{a41},\end{equation}
where $\psi_i(E)$ ($\phi_i(E)$) indicates the particle (hole) wavefunction
of energy $E$ on site $i$ and $j$ sums over the neighbours of $i$.
Since only scaling behaviour near a critical point is of interest,
we examine the simplest possible model
of a system with no normal disorder, but a
spatially fluctuating order parameter, obtained by
choosing $\epsilon_i$ equal to a constant $\epsilon_0$ for all sites $i$ and
to set the energy scale, choose $\gamma =1$.
Two models of disorder will be examined. In model 1, (which preserves time
reversal symmetry), we choose $\Delta_i=\Delta_0[1+\delta\Delta_i]$
and in model 2, (which breaks time reversal symmetry), we choose
$\Delta_i=\Delta_0[(1+\delta\Delta_i)+\imath([1+\delta\Delta'_i]$,
where $\delta\Delta_i$ and $\delta\Delta'_i$ are
random numbers uniformly distributed
between $-\delta\Delta$ and $+\delta\Delta$. In what follows, 
we choose $\epsilon_0=0$.

For each model, we compute the transfer matrix $T$ for a long strip
($d=$ 2) and a long bar ($d=$ 3) of length $L$ sites 
and cross-section $M^{d-1}$ sites, respectively and identify
the localisation length $\xi_M$ with
the inverse of the corresponding smallest Lyapunov exponent.
The results are, of course, sensitive to the chosen energy $E$
and since, in the absence of disorder (ie $\delta\Delta=0$),
there exists an energy-gap at $E=0$, the usual choice of $E=0$
adopted in the absence of superconductivity is inappropriate.
As a guide to a reasonable choice of $E$, we consider the related problem
of a system with normal disorder but with a uniform order parameter.
In this case  $\epsilon_i$ is chosen randomly from a uniform
probability distribution but $\Delta_i=\Delta_0$, for all $i$.
As noted in \cite{20} if $\psi^0_i(E_0)$ is a solution of
   
\par
\vspace{.4in}
\centerline{\psfig{figure=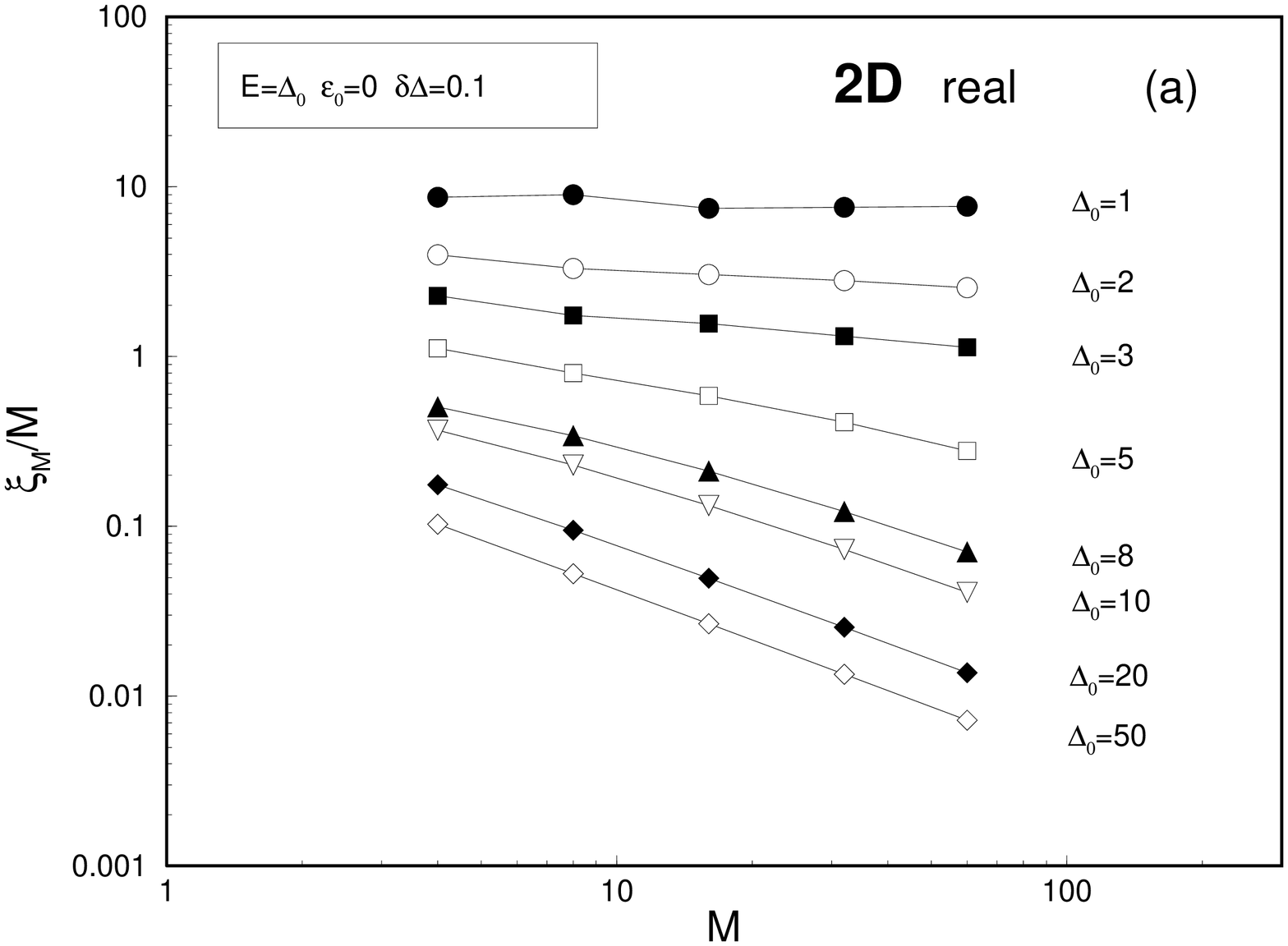,height=2.3in,angle=-90}}
\vspace{.2in}
\centerline{\psfig{figure=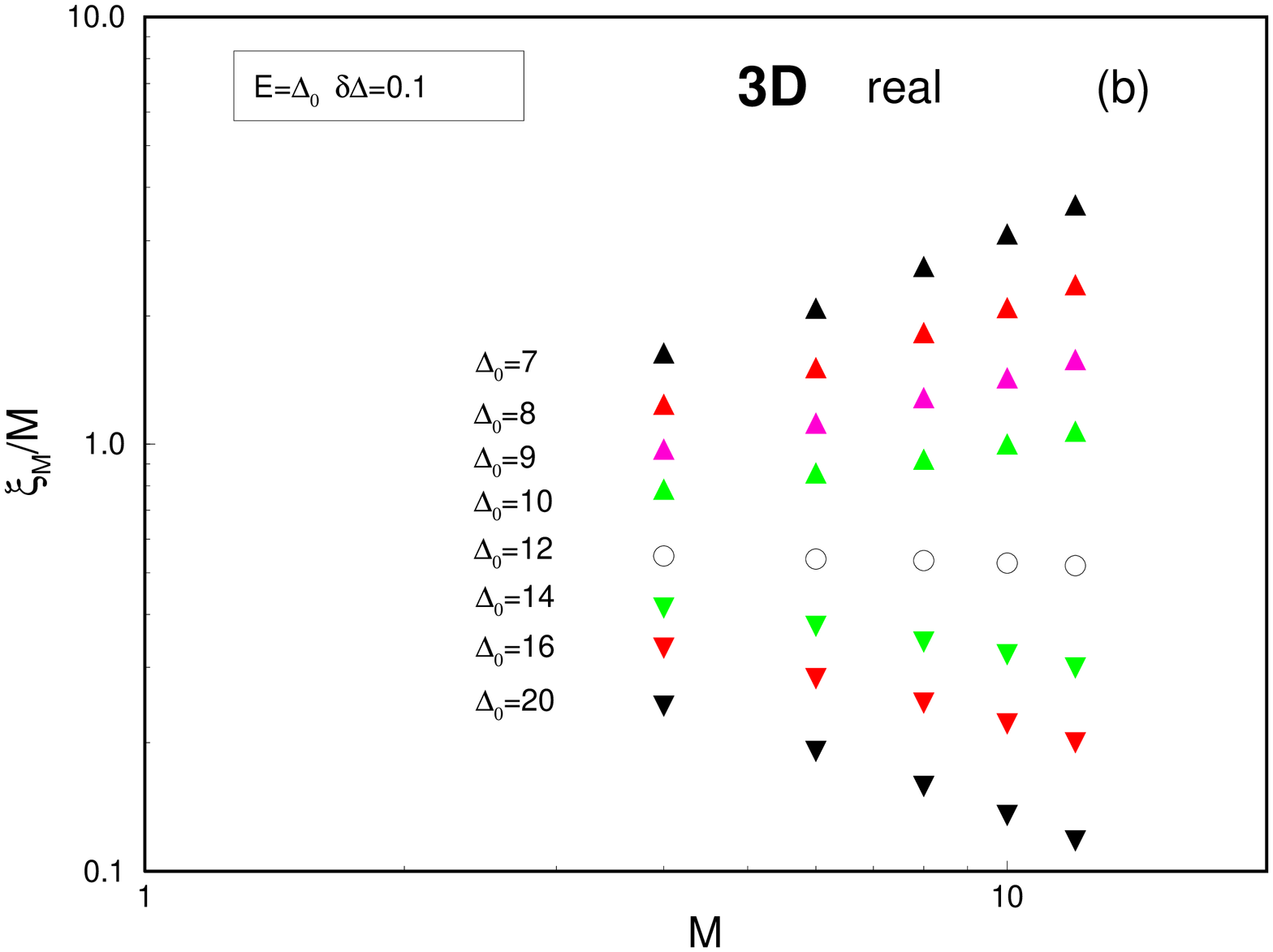,height=2.3in,angle=-90}}
{\footnotesize{{\bf FIG. 1.}
(a) The $\xi_M/M$ plotted as a function of the finite width 
$M$ in two dimensions for $E=\Delta_0$, $\epsilon_0=0$, $\delta\Delta=0.1$, 
and various values of $\Delta_0$.
(b) As in (a) but in three dimensions where a critical point is indicated.
}}
\par
\vspace{.4in}
   
\noindent
the normal-state 
Schrodinger equation, namely
$\epsilon_i \psi^0_i(E_0)-\sum_j \psi^0_{j}(E_0)=E_0\psi^0_i(E_0)$,
then $\psi_i(E)$ and $\phi_i(E)$ are each proportional to
$\psi^0_i(E_0)$, where $E=\sqrt{[E_0^2+\vert\Delta_0\vert^2]}$.
This means that if in the absence of superconductivity
a state at energy $E_0$ is localised by normal disorder, then
in the presence of a uniform order parameter $\Delta_0$, quasi-particle
states at energy $E$ are localised with the same localisation length.
As a consequence all critical properties are unchanged, provided
$E_0$ is replaced by $E$. In the normal state problem the least
localised states occur at $E_0=0$ and therefore in the presence of
normal disorder and a uniform superconducting order parameter these states
occur at $E=\vert\Delta_0\vert$.
Of course,  in what follows we are interested in the opposite limit
of a spatially fluctuating order parameter with no normal disorder.
Nevertheless, guided by the above observation
we choose $E=<\vert\Delta_i\vert>$,
where $<\vert\Delta_i\vert>$ is the ensemble averaged
order parameter, which gives $E=\Delta_0$ for model 1 
and $E=\sqrt{2}\Delta_0$ for model 2.

The raw data for $\xi_M/M$ versus $M$, 
for model 1 with $E=\Delta_0$, 
$\epsilon_0=0$ and $\delta\Delta=0.1$, 
are shown in figures 1(a) and (b) for two and
three dimensions,  respectively. 
The strength of disorder in the order
parameter is $W=2 \Delta_0\delta\Delta$, whose critical value is denoted
$W_c$, and is varied
by changing $\Delta_0$, with fixed $\delta\Delta$.
In two dimensions $\xi_M/M$ decreases with
increasing $M$ indicating that all states are localised
with $W_c=0$, whereas
in three dimensions there is a cross-over from 
localised to extended behaviour
at around $\Delta_0\approx 12$ which for the adopted value of
$\delta\Delta=0.1$ corresponds to $W_c\approx 2.4$.

To quantify the critical behaviour in 3 dimensions, we
linearize  the data  about $W_c$ 
by writing $\log (\xi_M/M)=\alpha_M + \beta_M \log W$
and obtain the  coefficients $\alpha_M$ and $\beta_M$ for various $M$.
In terms of the fixed point values $\log (\xi_M/M)_c$
and $\log W_c$, we note that
$\alpha_M=\log(\xi_M/M)_c- \beta_M \log W_c$. Thus, a graph of
$\alpha_M$ versus $\beta_M$
yields
$\log(\xi_M/M)_c$, $-\log W_c$ and hence the critical disorder
$W_c$.
The critical exponent $\nu$
for the divergence of the localization length $\xi$ of the infinite system
is obtained
by substituting $\alpha_M$ into the first linear relation,
which yields
$\log (\xi_M/M)=\log(\xi_M/M)_c + \beta_M\log(W/W_c)$.
Moreover, near the critical point 
$\log(W/W_c) \sim (W-W_c)/W_c$ and $\xi\sim |W-W_c|^{-\nu}$, so that 
$\log (\xi_M/M)=\log(\xi_M/M)_c\pm \xi^{-1/\nu}\beta_M$,
where the + (-) sign refers to $W>W_c$ $(W<W_c)$. The 
finite size scaling requirement 
$\xi_M/M=f(\xi/M)$ immediately implies $\beta_M\sim M^{1/\nu}$,
which permits the computation of the exponent $\nu$. 

Figure 2 shows a graph of $\log (\xi_M/M)$ 
versus $\log \Delta_0$,
from which $\alpha_M$ and $\beta_M$ for the chosen
widths $M$ can be extracted. 
The top-right insert shows the resulting plot 
of $\alpha_M$ versus $\beta_M$
whose slope is $-\log W_c$ and the corresponding 
intercept is $\log(\xi_M/M)_c$.
This yields $W_c=2.36\pm 0.04$ which corresponds to
$\Delta_0 =11.73\pm 0.12$ and $(\xi_M/M)_c=0.58\pm 0.02$. 
The lower-left
insert shows $\log \beta_M$ versus $\log M$ whose slope 
yields the critical exponent $\nu = 1.64 \pm0.06$.
   
\par
\centerline{\psfig{figure=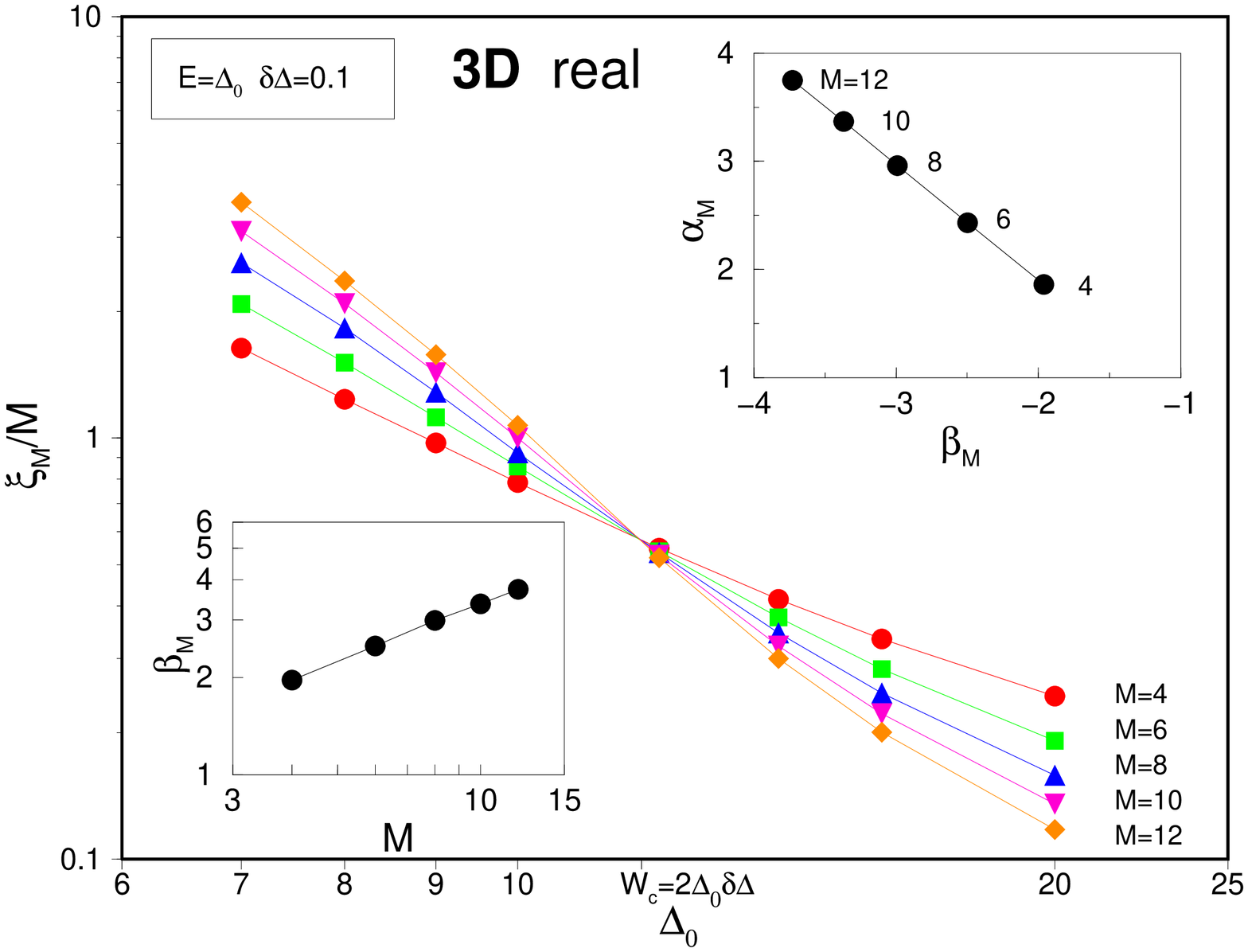,height=2.3in,angle=-90}}
{\footnotesize{{\bf FIG. 2.}
Log-log plot of $\xi_M/M$ versus $\Delta_0$ where 
the intersection defines $W_c$ and $(\xi_M/M)_c$.
The upper-right insert shows the coefficients $\alpha_M$ versus $\beta_M$
whose slope is $-\log W_c$. 
The lower-left insert shows a log-log plot of $\beta_M$ versus $M$
which yields the value for the exponent $\nu = 1.64 \pm0.06$.
}}
\par
\vspace{.4in}
   
\noindent
For model 2, where time reversal invariance is broken due to
the presence of a complex order parameter,
all states are localized in $d=$2. In contrast,
figure 3(a) shows the corresponding plots of $\xi_M/M$ versus $M$ 
in 3 dimensions which clearly show
a cross-over from extended to localised behaviour.
Results from a more accurate calculation are presented in figure 3(b),
where the upper-right figure yields
$W_c=5.57\pm 0.12$ and $(\xi_M/M)_c=0.58\pm 0.02.$ 
The lower-left
insert shows $\log \beta_M$ versus $\log M$, the slope of which
leads to the value for the exponent $\nu = 1.69 \pm 0.06$.
The errors in the calculation of $\xi_M /M$ are monitored as a function
of length $L$ and chosen to be less
than about 0.01 by taking long strips of lengths $L=250000$ and 
bars of $L=200000$ ($L=50000$) for the real (complex) case.
The errors for $W_c$ and $\nu$ are estimated from
the corresponding least-square fits.
We have also repeated our calculations 
by taking points closer to the critical value $W_c$, where the 
above analysis holds, with no significant change of our results.

The first important feature of the above calculation is the
unambigous prediction of  superconductivity-induced
quasi-particle localization
in $d=$2 and the presence of a mobility edge in $d=$3.
Localization arises from fluctuations in the
superconducting order parameter alone, without the need for
additional normal disorder. A second key result 
is the observation that for both models we find
$(\xi_M/M)_c\sim 0.58$ and 
$\nu\sim 1.6$, which are remarkably close to the values reported for
normal $d=$3 real systems \cite{21}, and  
also consistent with reported data
for ordinary disordered critical systems 
with and without time-reversal invariance \cite{22,23}. 
Recently, sligthly different scaling behavior is obtained 
with and without time-reversal by an alternative data analysis
based on polynomial fits \cite{24}. Our study in $d=$ 3 cannot 
distinguish such small difference if it exists.

\par
\vspace{.4in}
\centerline{\psfig{figure=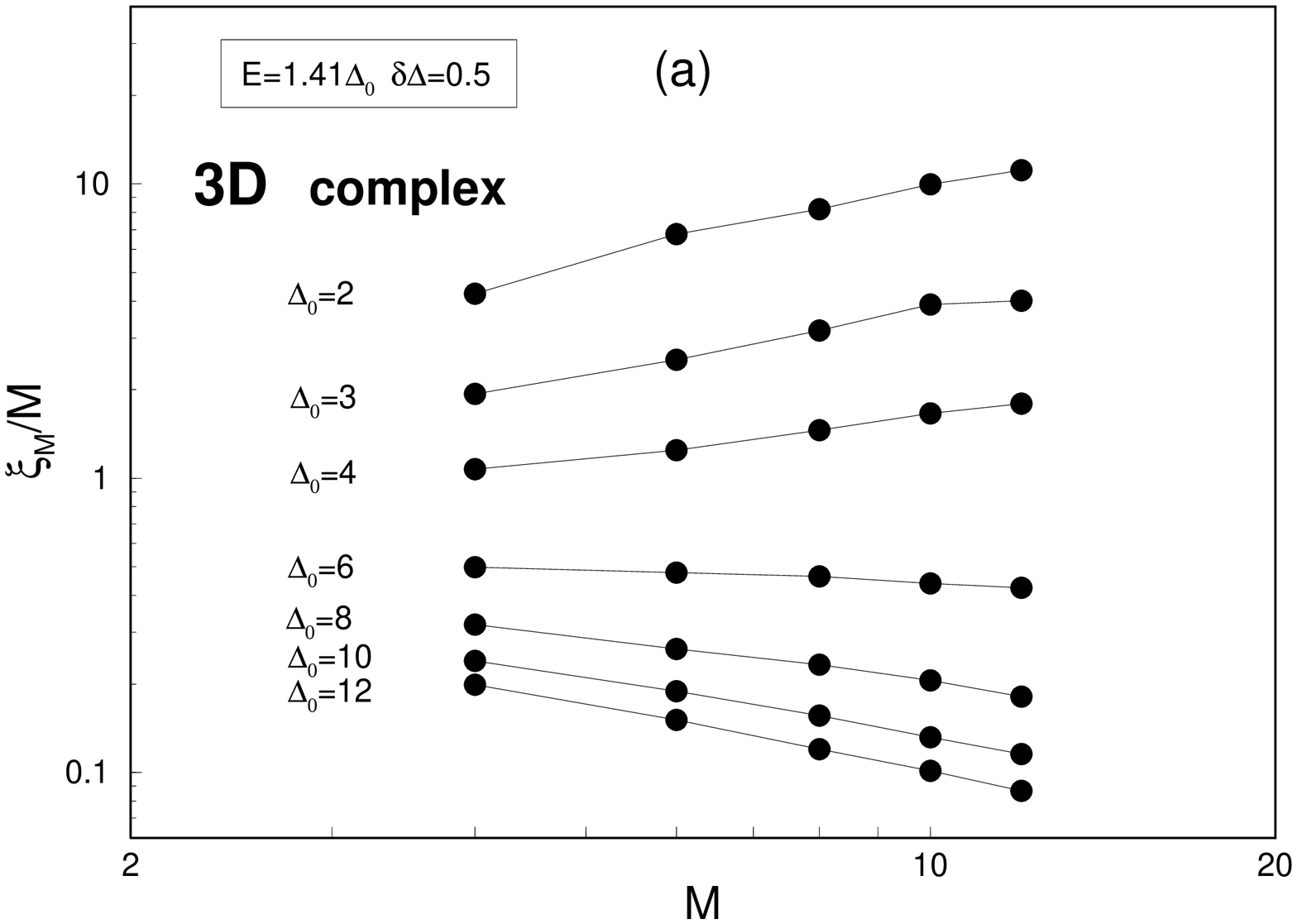,height=2.3in,angle=-90}}
\vspace{.2in}
\centerline{\psfig{figure=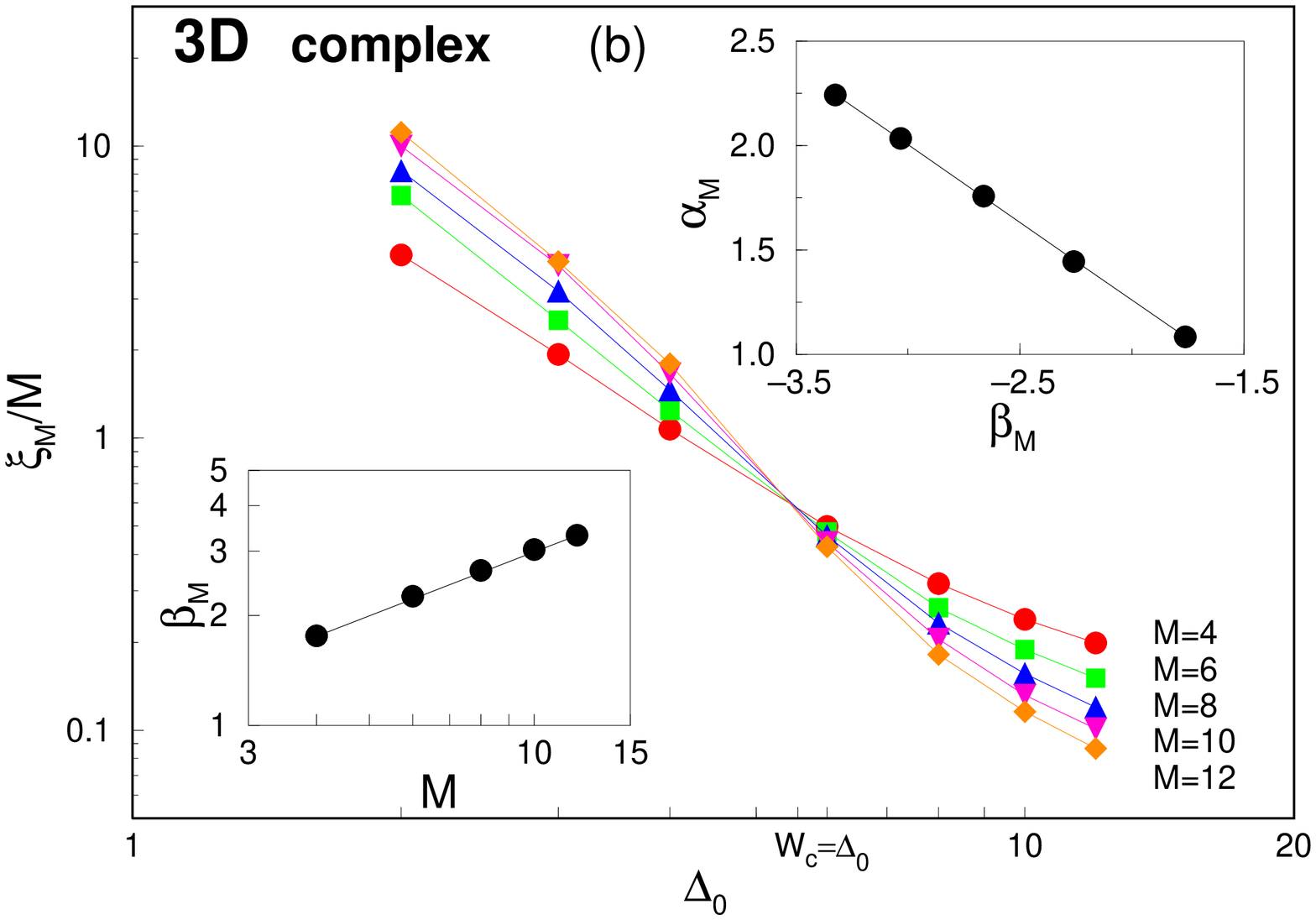,height=2.3in,angle=-90}}
{\footnotesize{{\bf FIG. 3.}
(a) Log-log plot  of $\xi_M/M$ versus $M$ for various values 
of $\Delta_0$ when time reversal symmetry is broken (model 2)
which shows a cross-over from extended to localised states.
(b) Log-log plot of $\xi_M/M$ versus $\Delta_0$ for model 2
where the intersections define $W_c$ and $(\xi_M/M)_c$.
The upper-right insert shows the coefficinets $\alpha_M$ versus $\beta_M$
whose slope is $-\log W_c$. 
The lower-left insert shows a log-log plot of $\beta_M$ versus $M$
which yields the value for the exponent $\nu = 1.69 \pm0.06$.
}}
\par
\vspace{.4in}
   
From an experimental point of view, it is worth noting
that the absence of quasi-particle diffusion does not imply 
the vanishing of the electrical
conductance, because Andreev scattering does not conserve quasi-particle
charge. It does, however, imply a vanishing of the electronic
contribution to thermal transport from certain states above the gap.
In a clean superconductor at a finite
temperature $T$, this varies as $\exp(-\Delta/k_bT)$, where $\Delta$ is
the bulk energy gap. In contrast, in the presence
of a fluctuation-induced quasi-particle mobility edge $E_c$, 
this will be replaced by $\exp(-E_c/k_bT)$. Thus, for example, the
melting of a flux lattice in a high temperature superconductor
should be accompanied by  an exponential change in the electronic
contibution to the thermal conductance.

This work was supported by a TMR of the EU 
and for D.E.K. and S.N.E. partially by the Greek 
Secretariat of Science and Technology, 
who  are also grateful for the 
hospitality at Lancaster where most of the work 
was carried out.

\end{document}